\documentclass[11pt,twoside]{article}
\usepackage{asp2004}
\usepackage{psfig}
\usepackage{epsf}
\usepackage{graphics}
\usepackage{lscape}
\def\edcomment#1{\iffalse\marginpar{\raggedright\sl#1\/}\else\relax\fi}
\reversemarginpar

\begin{document}
\title{Searching for Early Ionization with the Primeval Structure Telescope}
\author{Jeffrey B. Peterson}
\affil{Carnegie Mellon University, Department of Physics, 5000 Forbes Ave.,
Pittsburgh PA 15217 USA, email:jbp@cmu.edu}

\author{Ue-Li Pen}
\affil{
Canadian Institute for Theoretical Astrophysics, 
60 St. George St., Toronto, M5S 3H8, Canada, email:pen@cita.utoronto.ca}

\author{Xiang-Ping Wu}
\affil{National Astronomical Observatories, Chinese Academy of Sciences, 
20A Datun Road, Beijing 100012, P. R. China, email: wxp@bao.ac.cn}

\begin{abstract}

The Primeval Structure Telescope (PaST), will be used search for and study the
era the of the first luminous objects, the epoch of reionization. 
The first stars ionized the gas around them producing a pattern of ionization that reflects the
large scale density structure present at the time. The PaST array will be used in an attempt to sense and study this ionization,
by mapping the brightness of 21-cm neutral hydrogen emission at redshifts from 6 to 25. This emission disappears on ionization, 
allowing the study of large scale structure and of star formation at this very early epoch. 
The 10,000
antenna PaST array will be used to image ionized structures by creating
1 million pixel images of the sky.  The  angular scales of the images to
be produced span from 5 arc-minutes 
to 10 degrees. The array is currently under construction and over 2000 antennas have been installed.

\end{abstract}
\thispagestyle{plain}

\section{Introduction}

The Primeval Structure Telescope (PaST)(Peterson, Pen \&  Wu 2004, Pen, Wu \& Peterson, 2004) 
is designed to study the history of the Universe from age 100 million to 1 billion years. 
PaST will create three dimensional images sensitive to 21-cm neutral hydrogen emission at
high redshifts, which we will use to examine the ionization state of the early Inter Galactic Medium.  The first stars
ionized broad regions of the IGM, creating high contrast 21-cm sky structure at few-arcminute angular scales.  We will
search for that structure, to study early star formation, large scale structure and the world model.

When completed PaST will have about 30,000 square meters of collecting area and will operate in 
the neglected VHF band from about 50 to 200 MHz. PaST will be used to observe a single 100 square degree field for many months, 
creating multi-octave sky images far deeper than have ever before been made.

\section{Early Ionization}

The most exciting and unexpected result to come from the recent
Wilkinson Microwave Anisotropy Probe (WMAP,
Kogut et al. 2003) is the hint that the Universe may
have undergone the transition from neutral to fully ionized 
much earlier than had been previously thought.
Before WMAP the period between the emission of microwave background photons and
the emergence of the highest redshift galaxies and quasars ($z \sim 6$) yet
seen (e.g., Fan et al. 2003)
was thought by many to be dark. But it is now widely recognized that this epoch, beyond the
current high redshift observational frontier, is a crucial period in 
the history of the Universe. This so-called ``dark age'' is probably far from dark.
As the first sources of light switched on, 
they made the intergalactic gas   
luminous.  PaST radio observations can be used 
to map this material, as it undergoes a dramatic and as yet unseen transformation.

The emission we are searching for is from the
$1420.4/(1+z)$ MHz spin-flip transition
of neutral hydrogen (HI). The Universe at these early times
was filled both with Cosmic Microwave Background (CMB)
photons and with neutral hydrogen in
the Intergalactic Medium (IGM). In equilibrium, 
the HI spin temperature and the CMB temperature would be the same, and no
21-cm emission would be visible on the sky. However, as soon as a few of the
first stars turned on, the photons they produced were sufficient to excite
the hydrogen atoms from the ground state to the 2p state. The atoms returned to the
ground state, but in either the hyperfine singlet or triplet state. In this way,
the spin temperature became coupled to the gas kinetic temperature
(Wouthuysen 1952, Field 1959) and emission from the IGM should still be visible
against the CMB today.

\begin{figure}[ht!]
\plotone{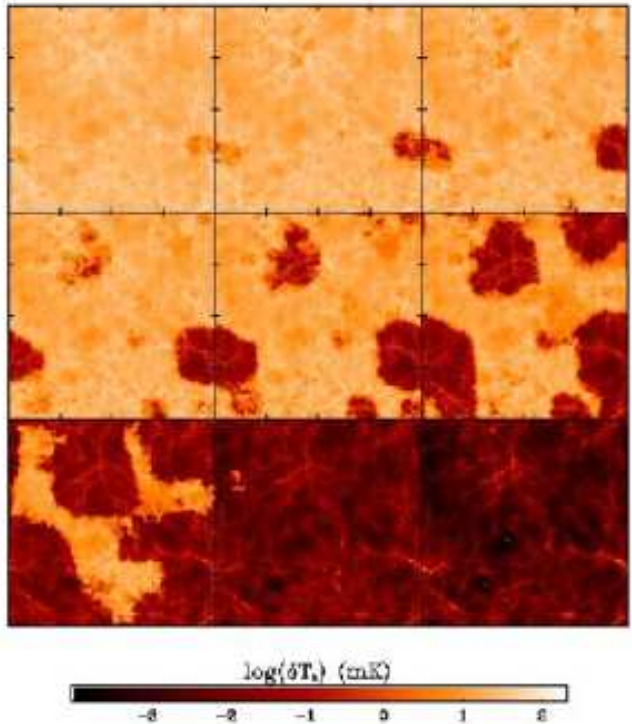}
\caption{{\bf Simulated VHF sky images. } The nine
panels show simulated sky brightness in redshifted 21-cm emission,
separated in time, beginning in the upper left. Already in the
first time slice the densest region, at the right of the panel, is
losing brightness as the first stars are beginning to ionize the
hydrogen gas. By the final panel the entire sky is dim--the IGM is
ionized. Note the high contrast in the intermediate images. The
typical patch size is expected to be around 5 arcminutes. In this particular model, the
total redshift range of the transition corresponds to about 6
MHz in the received spectrum. Plot provided by S. Furlanetto}
\end{figure}

Early pioneering work on this 21-cm wavelength radiation as
a probe of the early universe
was carried out by Sunyaev and Zeldovich (1972). Later predictions
in the context of more modern models of structure formation were made by
authors including Scott and Rees (1990), Madau, Meiksin \& Rees (1997) 
and Gnedin and Ostriker (1997). Recently,
there has been a flood of theoretical work, much of it motivated
by the WMAP results. 21-cm emission from the IGM in cosmological
hydrodynamic simulations has been studied by Gnedin \& Shaver (2003),
Furlanetto, Sokasian \& Hernquist (2004) and Ciardi \& Madau (2003).

These authors find that the evolution of HI emission is
more abrupt than is seen in the large-scale distributions of galaxies, the
former being driven by the advance of ionization fronts and
the latter by gravity. In Figure 1 (taken, with permission
from Furlanetto, Sokasian \& Hernquist 2004), we show the brightness 
temperature variations on the plane of the sky at a sequence of times. 
By scanning through frequencies, these different stages can be sorted by redshift, 
making 3-D tomography of the IGM possible.
In the early panels (top left), the IGM is luminous, being
completely neutral. At this stage, the image is truly representative of the
hydrogen density structure of  IGM.

In the second and later panels,
ionizing sources of radiation have switched on, as star formation progresses from
high-density regions to those of lower density.
Once a region is ionized, ultraviolet radiation propagates through it, 
and further ionization occurs at the margin. This causes 
large abrupt-edged patches of the gas
to go dark in 21-cm emission,
providing a visual tracer of the ionization state of the IGM. 21-cm sky structure contains
a wealth of information on the nature of the first sources of radiation, including their 
luminosity, space density and radiation spectra.  However, even though the details
in this image convey a wealth of information, on large scales, the underlying power spectrum
reflects the power spectrum of the pre-existing density structure.  This means
we know where in angular scale to search first for the ionization signal--there should
be substantial power at arcminute scales.

Before the WMAP results, it was widely thought that the partial ionization state 
occurred relatively late, at around redshift
$z=7$.  This was motivated by the increase in the optical depth to hydrogen
Lyman-alpha absorption seen in quasar spectra close to this
epoch (Fan et al. 2002, Becker et al. 2005). The recent evidence of a large variance between
sightlines to high redshift quasars (e.g. Oh \& Furlanetto 2004)
indicates that the ionization transition is complex.
Many scenarios are being advanced to explain the combination of CMB and
quasar results, including
the possibility of two separate periods of reionization (Cen 2003,
Wiythe and Loeb 2003).
Direct observations from PaST will enable the process of reionization
to be seen as it happens, so that questions about the nature
of the first sources of light and the stages of reionization can
be properly addressed. 

\begin{figure}[ht!]
\plotone{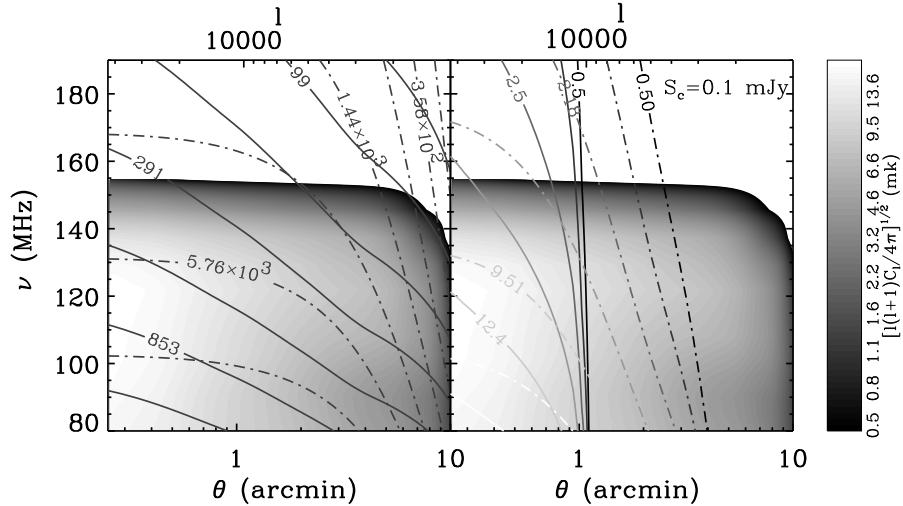}
\caption{
{\bf VHF Extragalactic Foreground. } The 
shaded area is the 21-cm signal (power spectrum) from high
redshift for a rather gradual early ionization model.
The black lines are the contours (labeled with the respective values in
mK) for a predicted sky fluctuations due to radio-galaxies. In the left panel no source
subtraction has been made. The foreground signal overwhelms
the 21-cm signal PaST seeks.
In right hand panel, sources with flux greater than 0.1 mJy were removed.
The amplitude of the radio galaxy signal is significantly
reduced and in particular the power in the smaller scales
drops to well below the Early Ionization signal. To accomplish this foreground removal the PaST team
will need to understand the PaST instrument response quite well. The radio galaxy
sources themselves will provide this information. Plot by T. Di Matteo. (Di Matteo, Ciardi \& Miniati 2005)}

\end{figure}

\subsection{Future directions}

There are many tests of cosmology
that can be envisaged using the 21-cm signal. For example, because frequency 
information extends the observable domain into the third dimension, it 
will be possible to use the 
difference between angular and redshift clustering to measure 
cosmic geometry via the Alcock-Paczynski test (Nusser 2004).

In general, the 21-cm signal contains information on 
clustering of matter on much smaller scales than other 
cosmological probes of structure
such as galaxy redshift surveys, the Lyman-alpha forest, or the CMB. 
This means that with high redshift 21-cm data theories of dark matter and 
inflation can be tested in a new small-scale regime, where model predictions of the power spectrum
of fluctuations can differ dramatically (Profumo et al. 2004,
Kamionkowski \& Liddle 2000).

An ambitious future program would be an attempt to detect
neutral hydrogen absorbing the CMB in the epoch before the
formation of the first sources of light (Loeb and Zaldarriaga 2004).
In the redshift range from
$z=200-30$, the temperature of the gas in the Universe is below the CMB
temperature. Then, collisional excitation coupled the spin temperature to the
gas kinetic temperature, and as a result the cool gas can be seen in absorption
against the CMB. At these early times, no star formation is expected
to disrupt the simple gravitational processes at work and theoretical
predictions are easy to make. The amount of potential information available, is
orders of magnitude larger than for any other cosmological probe.
The PaST program will let us take preliminary steps in this exciting direction.

The hydrogen absorption signal peaks at around $z=70$ which corresponds to 20 MHz. This is the
short wave (a.k.a. HF) band: radio broadcasts here propagate across the entire mid-latitude region of the planet.
An antarctic site may be
useful for short wave observations, since HF ionospheric over-the-horizon propagation 
shuts down in the Polar winter. However auroral activity may make
the ionosphere turbulent in the polar regions, creating poor ``seeing''.  
In addition to the construction of the PaST array in China we are carrying out a test of the South Pole
as a possible future site for VHF/HF astronomy.

\section{Foreground Removal}

In order to separate the 21-cm emission from foregrounds, PaST will image the sky brightness at many frequencies.
While it may be possible to use simple statistical measures of sky brightness fluctuations as a function 
of redshift to differentiate between reionization scenarios
(Ciardi \& Madau 2003, Furlanetto, Sokasian \& Hernquist 2004), with PaST
we have decided to build in sufficient sensitivity to create high
signal to noise images for each few-megahertz of span of bandwidth.

The main complication involved in extracting the 21-cm cosmological
information from a VHF sky image comes from foreground sources
of radiation. The point source foreground fluctuation level
exceeds the 21-cm sky brightness structure by more than a factor 100
(Di Matteo et al. 2002, Oh \& Mack 2003) as shown in figure 2.
We expect that the foregrounds can be separated from the 21-cm signal
because all postulated contaminants have smooth spectra, whereas the
cosmological signal fluctuates sharply in frequency(e.g., Di Matteo  et al. 2002, Gnedin \& Shaver 2003). 
For example, the transition between the very different images in panels 6 and 7
in Figure 1 takes place over only 6 MHz in the received spectrum. Over this small frequency range
the foreground fluxes vary only slightly.

\begin{figure}[ht!]
   \plotone{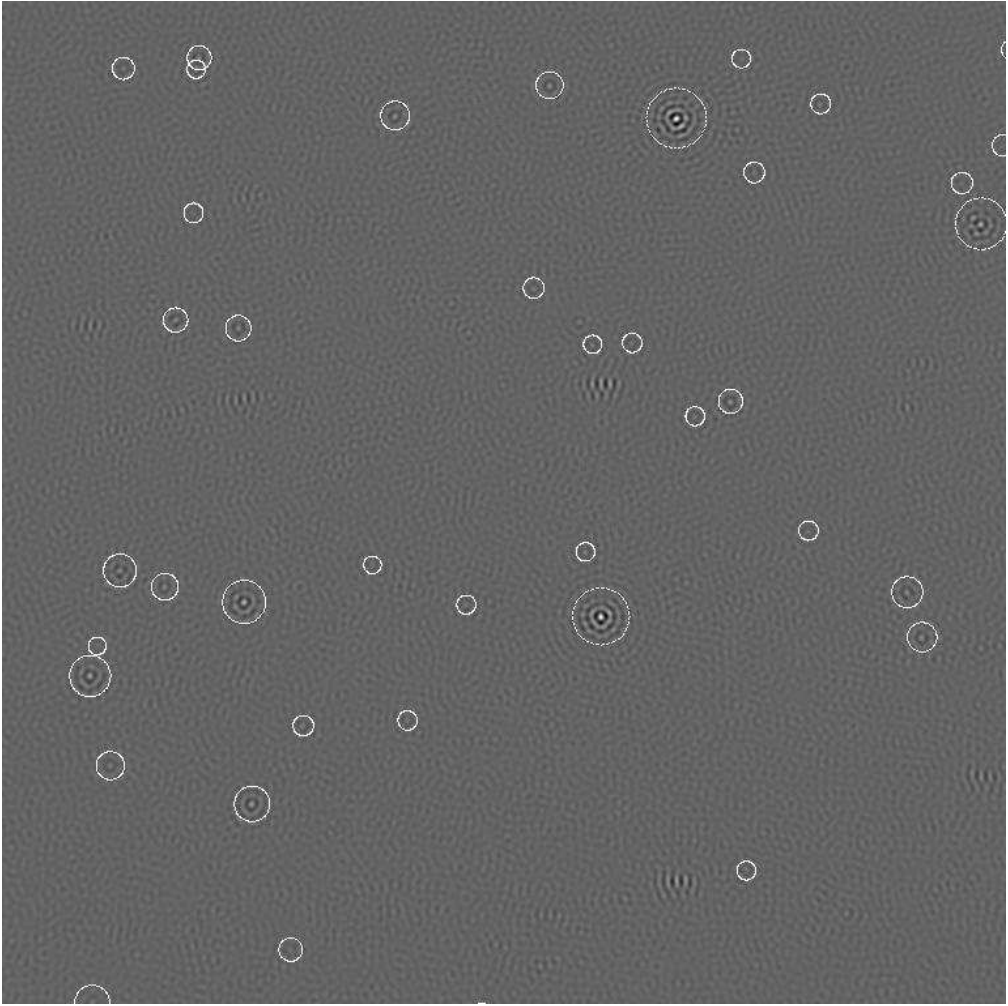}
   \caption{{\bf Sky Image Created with a PaST Prototype}
This image was reconstructed using a single 1088m east-west baseline.  The
displayed field of view is 24 degrees.  Frequency range is 62-88
MHz. Overlaid on the gray scale
map from the PaST prototype are 8C sources with flux greater than 10
Jansky shown as circles.  The area
of each circle is proportional to the 8C flux.
This image was created using 12 hours of data and 14 antennas.}
\end{figure}

\section{Secondary Science}

Although the study of early ionization is the primary science project for PaST, we will have an unprecedented 
combination of collecting area, frequency coverage and observation duration. We therefore must remain alert to
possible new astronomical sources.

Gamma Ray Burst sources probably also emit prompt electromagnetic pulses. Since the EMP and gamma flux are likely beamed 
in different directions it
 makes sense to mount not just a directed search for EMP-GRB coincidences
but also a blind search for EMP events.

If a sample of EMP-gamma coincidences reveals itself, that data set may be very useful in nailing down the
cosmological world model--especially the equation of state of dark energy. We know the IGM is highly ionized since redshift
6 and we have a reasonable estimate of the electron density.  So, the plasma delay, measured via the dispersion of the EMP, 
provides a measure of the physical path length to the source.  Combined with redshift information, a set of EMP 
dispersion measurements would allow the history of the expansion to be traced. In principle, one might be able to directly witness 
the shift from deceleration of the early universe to the acceleration that seems to be occurring at this epoch.  

But these goals run far ahead of observation. There are so far no convincing EMP candidates. 
Data we have obtained on-site, with prototypes of PaST, indeed show short bursts of VHF energy, as shown in figure 3.  
We believe these events to be terrestrial, since they lack dispersion. They are
likely due to over-the-horizon propagation of VHF signals relayed to Ulastai via ionized meteor trails. 
We know that we will need to catalog and cut these events, so our standard data analysis software already 
has pulse detection filters.
By de-dispersing the spectrum, our pulse finding algorithms
can also identify astronomical EMP candidates.

Another proposed instrument, the MWA-demonstrator, has as a science goal the search for EMP candidates.  The two instruments will
operate in very different niches.  PaST will have a much larger collecting area than MWA-demo, but while MWA-demo 
can place a beam at any 
point on the sky, PaST will always stare at the
same field.  This makes MWA-demo 
particularly useful for gamma-coincidence searches, since GRBs occur at random across the sky. 
Meanwhile, PaST can be used to do a deep blind EMP search. 
Of course, it also makes sense to examine the PaST data for EMP signals coincident with
GRB events that occur in our field of view.

\begin{figure}[ht!]
   \plotone{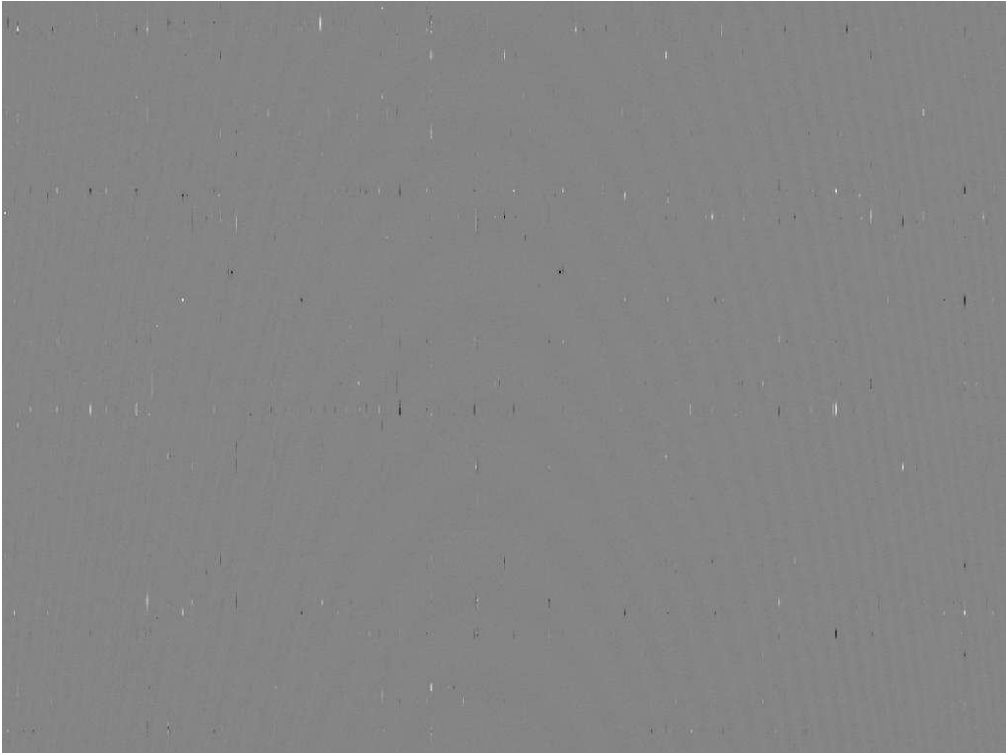}
   \caption{
{\bf Radio Frequency Interference.} Reflected RF signals, scattered by meteor trails, are seen in the raw correlator output.  
Time runs to the right, and a total of 
4 hours are shown.  Frequency increases upward,
covering 75-82 MHz.  Only the real part of the correlation is shown,
and the sign of the signal (black vs white) relates to the phase
due to the direction of 
the source.  Time resolution is 20 seconds.  The occasional flashes are believed to be 
due to ionized meteor trails creating over-the-horizon
propagation paths. This carries RF signals to Ulastai from nearby cities.
The nearest city is 200 km away so, between meteor events, the site is extremely quiet.} 
\end{figure}

\begin{center}
\begin{table}[t]
\caption{\bf Specifications of Planned PaST Instrument. }
\begin{tabular}{|l|c|}
\hline
Parameter & Goal (at 150 MHz) \\
\hline
\hline
Total antennae                                  &10000\\
Array Elements                              &80\\
Effective area                                  &30,000 $m^2$\\
Sky brightness temperature                      &200 K\\
Instantaneous imaging field of view               &20 sq deg\\
Angular resolution              &3 arc minutes\\
Frequency range        &50 - 200 MHz\\
Instantaneous bandwidth 	&50 MHz\\
21-cm redshift range                &  6 - 27 \\
Frequency resolution                &4 kHz\\
Polarization     &Full Stokes\\
Comoving radial spatial resolution          &140 kpc\\
Comoving transverse spatial resolution      &0.6 Mpc\\
Brightness temperature sensitivity      &20 $mK/\sqrt{\rm day}$ \\
\hline
\end{tabular}
\end{table}
\end{center}

\section{PaST Instrument Description}

The instrument design is driven by our primary science goal: imaging ionization 
of large scale structure via 21-cm emission.
The redshift range of interest is 6 to 25, so the frequency range of the
telescope is about 200 MHz to 50 MHz. 
The frequency resolution is set at  4 kHz to remove terrestrial lines
during meteor scatter events.

The spatial scale range to be covered is determined by the need to avoid extragalactic foregrounds,
while still detecting the large scale structure that underlies the ionization pattern.
This is illustrated in figure 2. We selected 3 arcminutes at 100 MHz as the resolution of the telescope. 

The spatial resolution sets
the longest baseline at around 3 km.  This large size, combined with the relatively high contrast (20 mK)
of ionization structure, immediately points us to a dilute array as the appropriate technology for this telescope.

We want even UV coverage to faithfully generate images. 
Also, over some particularly interesting range of frequency
we may wish to re-analyze our data by synthesizing frequency-independent beams.  
That can be done using wavelength-scaled apodization
of the visibilities across the UV plane. 
This works best if there is dense, uniform UV coverage. To meet these requirements
we will use 80 polarization-sensitive interferometer elements, 
which we also call pods, each containing 127 log periodic antennas.

The required area of the elements depends on the system temperature, 
integration time, and required surface brightness
sensitivity.
The elements must have substantial antenna gain. 
We accomplish this by assembling the elements as phased arrays of log periodics.  
For this stage of the project we will use fixed phasing, with our main beam centered on the North Celestial
Pole.  If we find it necessary to do so we can later add switched delay lines within the phased arrays, allowing us
the option to beam-switch or to track fields not at the NCP.

A few technical details of the PaST system may be of interest. 
The PaST pods operate on solar-charged batteries and transmit signals via optical fiber.
This avoids possible couplings since no copper power, data or signal lines connect to the pods. Cable TV technology is used
within the pods: household signal splitters are used backward as signal combiners. Low noise amplifier designs for PaST
were contributed by amateur radio operators. The selected design has noise figure below 0.5 db.  
Although we tested loop antennas and log spirals,
we settled on a modified log periodic. These are similar to the millions of TV antennas manufactured in China so 
our antenna cost is about \$20 each.

Conveniently, the frequency range of the telescope  
is similar to the sampling rate available using a variety of off-the-shelf Analog to Digital 
Converters.  We use such ADCs to sample each RF signal directly into a PC computer.
The current PaST system has no mixers, no IF system and no local oscillators, although we may add these
if needed to improve instrument stability.
The FX correlator for PaST is organized as a network of PCs.
The current design incorporates aliased sampling at 100 M samples/s allowing
approximately 50 MHz of instantaneous observation bandwidth. 
All components of the correlator are commercial off-the-shelf items.

\begin{figure}[ht!]
\plotone{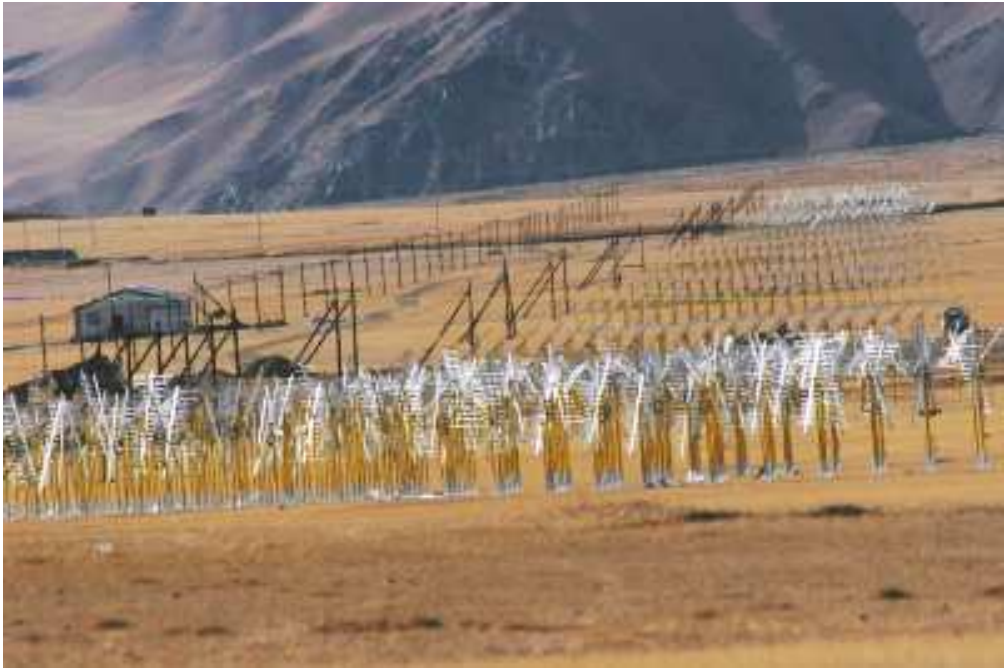}
\caption{{\bf Status of PaST, December 2004. } 
Over 2000 antennas have been installed and about half of these have have been aligned at the
North Celestial Pole. A 170 square meter building has been constructed. Power mains supply the site, as 
do copper telephone and internet connections. A 3 km optical fiber trunk to carry signals from the
phased arrays is in place.}

\end{figure}

\section{Prototype Tests}

During Spring and Summer 2004 the PaST team assembled, on-site, a proof of concept prototype array
consisting of four groups of seven log periodics. We used this prototype to 
evaluate a number of issues. We verified that aliased sampling, analog signal summing within the pod, and
intensity-modulated optical fiber signal relaying, all function in the field as anticipated. We also
field tested four different antenna designs, switched our transmission line impedance from 50 ohms to 75 ohms,
and tested a variety of low noise amplifiers, power combiners and bandpass filters.

We have used this prototype to study the process of phase calibration using astronomical sources.
The results of this study offer important information on the structure of the ionosphere. 
We find that we can easily establish phase
corrections across fields of ten degrees but that over radian-size fields simple phase correction models
are not sufficient. In the final PaST system the field of view will be about 10 degrees, well within the isoplanatic
patch of the ionosphere.  In contrast, both MWA and LOFAR will attempt all-sky observations.  They will need
complex, time-dependent, all-sky ionospheric modeling, while we will need at most gradient terms in our ionospheric model.
Figure 3 shows that even with a very primitive ionospheric model we can indeed achieve the resolution we need.
By restricting our field of view we have greatly simplified our data reduction task, allowing us to
focus on the celestial sphere, rather than the ionosphere.

The other important result of prototype testing is that we have shown the Ulastai site to be extremely quiet.
This is shown in figure 4.  Apart from occasional meteor scatter events, which are easily cut, the spectrum
is virtually a blank slate.

\section{Status of PaST}
A portion of the PaST array is shown in figure 5, as of December, 2004. 
More than 2500 log periodic antennas have been fabricated and transported to the Ulastai site
(E 86$^\circ$ 41', N 42$^\circ$ 56') and 2000 have been erected.
These antennas will be used to create a 25 pod array.  The 25-pods are not all identical.
Instead we are trying out a variety of phased array layouts and polarization properties. 
The East-West leg of the fiber optic signal relay system has been installed. Electrical power mains now service the
site, and a copper phone/data connection has been established. A 170 square meter building is now complete at the
site.

\acknowledgements{Funding for PaST in China has been provided by Natural Science Foundation of China (NSFC). Site testing at 
the South Pole is funded by the US National Science Foundation under grant OPP0342448. Tiziana Di Matteo and Steve Furlinetto provided
figures. JBP thanks Rupert Croft for useful discussions.}

\end{document}